\documentclass[aps,prx,twocolumn,showpacs,citeautoscript,superscriptaddress,longbibliography,notitlepage,usenames, dvipsnames]{revtex4-1}

\usepackage[english]{babel}
\usepackage[utf8x]{inputenc}
\usepackage[T1]{fontenc}
\usepackage{ulem}
\usepackage{graphicx}
\usepackage{pifont}
\newcommand{\cmark}{\ding{51}}%
\newcommand{\xmark}{\ding{55}}%

\usepackage[a4paper,top=3cm,bottom=2cm,left=3cm,right=3cm,marginparwidth=1.75cm]{geometry}

\usepackage{amsmath}
\usepackage[colorinlistoftodos]{todonotes}
\usepackage[colorlinks=true, allcolors=blue]{hyperref}

\begin{document}
\title{
Pseudo-electromagnetic fields in topological semimetals}

\author{Roni Ilan}
\thanks{All authors contributed equally to the preparation of this review}
\affiliation{Raymond and Beverly Sackler School of Physics and Astronomy, Tel-Aviv University, Tel-Aviv 69978, Israel}
\author{Adolfo G. Grushin}
\thanks{All authors contributed equally to the preparation of this review}
\affiliation{Univ. Grenoble Alpes, CNRS, Grenoble INP, Institut N\'eel, 38000, Grenoble, France}
\author{Dmitry I. Pikulin}
\thanks{All authors contributed equally to the preparation of this review}
\affiliation{Microsoft Quantum, Microsoft Station Q, University of California, Santa Barbara, California 93106-6105}

\begin{abstract}

Dirac and Weyl semimetals, materials where electrons behave as relativistic fermions, react to position- and time-dependent perturbations, such as strain, as if emergent electromagnetic fields were applied. Since they differ from external electromagnetic fields in their symmetries and phenomenology they are called pseudo-electromagnetic fields, and enable a simple and unified description of a variety of inhomogeneous systems involving topological semimetals. We review the different physical ways to create effective pseudo-fields, their observable consequences as well as their similarities and differences compared to electromagnetic fields. Among these difference is their effect on quantum anomalies, the absence of a classical symmetry in the quantum theory, which we revisit from a quantum field theory and a semiclassical viewpoint. We conclude with predicted observable signatures of the pseudo-fields and the nascent experimental status.

\end{abstract}

\maketitle

\section{Introduction}

In one of his 1964 Messenger Lectures at Cornell University, Richard P. Feynman said that "every theoretical physicist who is any good knows six or seven different theoretical representations for exactly the same physics"~\cite{Feynman1965}. 
This arguable statement gains support when different fields allow for complementary interpretations of reality, a blend that often leads to conceptual breakthroughs. High energy and condensed matter physics have been often interlaced in this way. Today, the three types of relativistic indivisible spin-1/2 fermions---Dirac, Weyl and Majorana fermions---are also quasiparticles in quantum matter
~\cite{Neto2009electronic,Hasan2010colloquium,Hasan2011three,Qi2011topological,Alicea2012,Beenakker2013,Lutchyn2018,Armitage2018}
notably in systems where global (topological) rather than microscopic properties classify quantum states~\cite{Schnyder2008, Qi2011}. 
\begin{figure}
    \includegraphics[width=\linewidth]{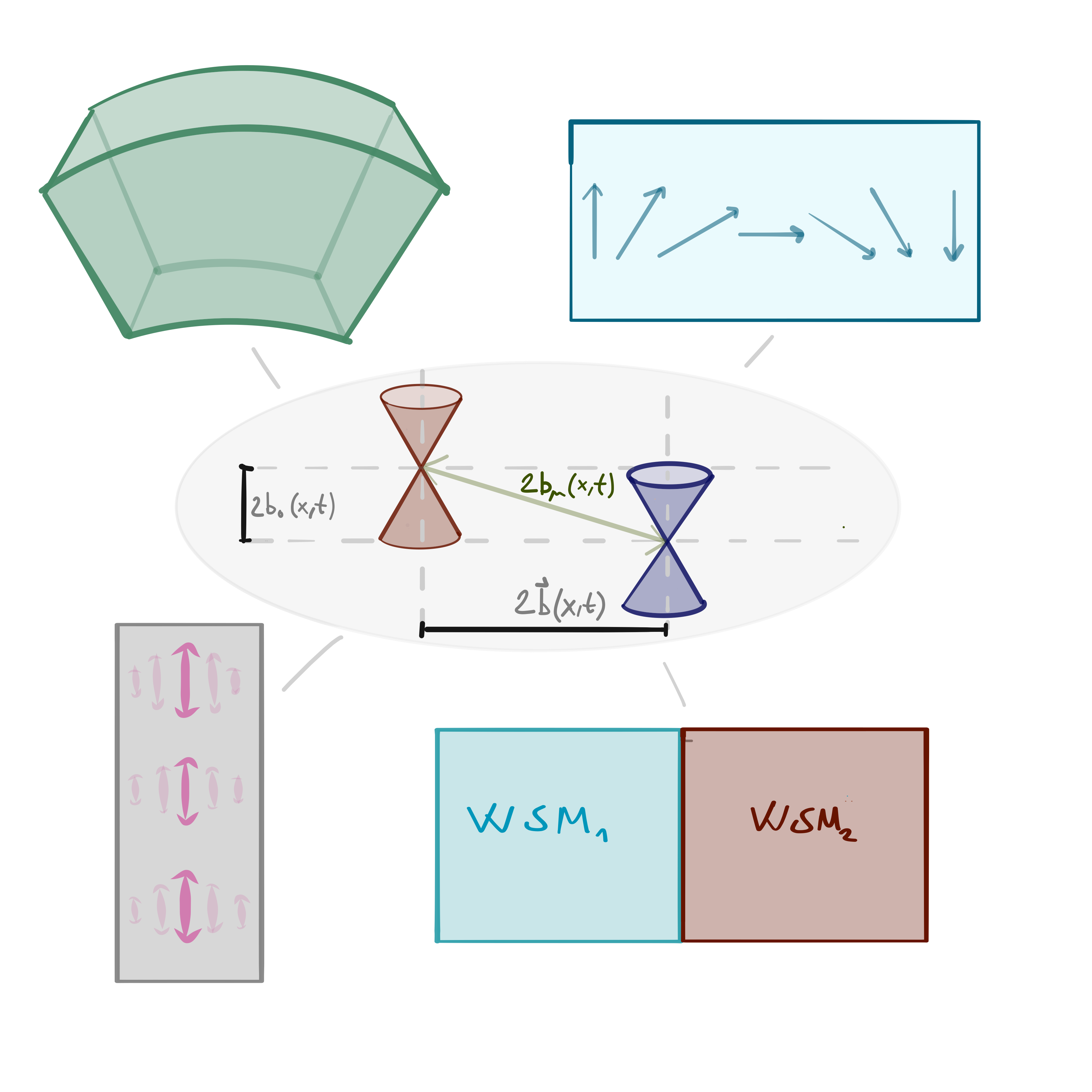}
    \caption{\textbf{Physical realizations of emergent pseudo-fields in Weyl semimetals.} The central panel shows two Weyl cones, of different chiralities (colors) that are separated in energy and momentum space by a scalar $b_0$ and a vector $\mathbf{b}$ allowed to vary in space-time. Derivatives of the combined (four-)vector $b_\mu(x,t)$ define emergent pseudo-electromagnetic fields which offer a simple unified viewpoint of different phenomena: a strained crystal, an inhomogeneous magnetization profile, a propagating sound wave, and an interface between two Weyl semimetals.}
    \label{fig:types}
\end{figure}

In condensed matter we are frequently required to describe systems where some underlying degrees of freedom are allowed to vary in space and time. Such is the case of a strained crystal, an inhomogeneous magnetization profile, a propagating sound wave, or an interface between two solids (see Fig.~\ref{fig:types}). When Hamiltonian parameters vary in space-time, the crystal symmetry is reduced allowing absent perturbations to emerge, such as Rashba spin-orbit coupling in ordinary semiconductors~\cite{Winkler2003}. The main idea in this review is that, so long as the quasiparticles experiencing space-time variations are Weyl fermions, their behaviour follows simple laws expressed in terms of emergent electromagnetic fields~\cite{Bevan1997,Liu2013,Shapourian2015, Cortijo2015}. In return, we can use this simple description to predict new phenomena in diverse contexts.

Weyl quasiparticles occur in Weyl semimetals where linearly dispersing non-degenerate bands touch at a single point in momentum space, a two-fold degeneracy that cannot be lifted~\cite{Yan2017, Armitage2018}. 
The dispersion resembles that of a photon, but Weyl quasiparticles have spin-1/2 and thus are described by an effective relativistic massless Dirac equation: the Weyl equation. 
Due to their effective relativistic symmetry and the absence of a gap, their Hamiltonian projects their momentum in a direction parallel or antiparallel to their spin.
The two cases are distinguished by a number, $\chi=\pm1$, the chirality. 
On a lattice, under very general assumptions like locality or unitarity, the number of quasiparticles of each chirality in equilibrium must be equal~\cite{Nielsen1981,Nielsen1981B}. 
Using Noether's theorem we may express this symmetry by a classical conservation law for the current that counts the difference between chiral or left moving ($\chi=1$) and anti-chiral or right moving ($\chi=-1$) fermions, the chiral current.

When quantaising the theory in the presence of electric and magnetic fields something unexpected happens: the quantum expectation value of the chiral current is not conserved, unlike its classical counterpart. A current that are conserved classically but not quantum mechanically are labeled anomalous~\cite{Bertlmann2000}. 

The non-conservation of the chiral current, the \textit{chiral anomaly}, has dramatic consequences in different fields of physics.
Weyl fermions mediate, as virtual particles, the decay process of the pion into two photons, which is off by an order of magnitude compared to experiments if the anomalous contribution is ignored! 
Weyl particles, however, have not been found  as elemental particles in nature, preventing us to probe the chiral anomaly directly. 
In contrast, Weyl quasiparticles do exist in Weyl semimetals and exhibit the chiral anomaly as an enhanced magneto-transport~\cite{Son2013}, confirmed experimentally~\cite{Yan2017,Liang2018}.

In analogy with other pseudo-relativistic fermions, such as emergent Dirac fermions in 2D materials~\cite{Guinea2010,Levy2010,amorim2016,Naumis2017}, it became clear that coupling Weyl quasiparticles to electromagnetic fields, which are blind to chirality, was not the only possibility. 
When the underlying parameters vary in space-time, electromagnetic fields that couple with opposite signs to opposite chiralities emerge in the low energy theory of Weyl quasiparticles.
Known as axial- or pseudo-fields, we may use them to reach a simple formulation of the different physical scenarios shown in Fig.~\ref{fig:types}. 
The goal of this review is to present and explain how pseudo-fields emerge in Weyl semimetals and the recent progress made in using them to describe Weyl systems. \\

 To understand the origin of axial-fields in Weyl semimetals consider  two linear three-dimensional band crossing of chirality $\chi=\pm1$ close to $\pm\textbf{b}$ and $\pm b_0$ in momentum and energy space respectively (see Fig.~\ref{fig:types}). Their low energy Hamiltonian reads
 \begin{equation}\label{eq:WeylH}
 H_\chi(\textbf{p})= v_F \chi(\textbf{p}+\chi\textbf{b})\cdot\sigma+\chi b_0.
 \end{equation}
For $\mathbf{b}=b_0=0$ this is the Weyl Hamiltonian of a pseudo-relativistic particle with momentum $\mathbf{p}$, where $\sigma$ are the spin-1/2 Pauli matrices. 
The vector $\textbf{b}$ enters the Hamiltonian in the same way a vector potential $\mathbf{A}$ would through a minimal substitution ($\mathbf{p}\to\mathbf{p}-e\mathbf{A}$), albeit sensitive to chirality. By analogy with $\mathbf{A}$, if $\textbf{b}$ has a smooth space dependence with non-vanishing curl, one may define
\begin{equation}
    \textbf{B}_5=\frac{1}{e}\nabla\times\textbf{b}\label{eq:B_5},
\end{equation} 
and interpret $\textbf{B}_5$ as an intrinsically emerging pseudo-magnetic field.  
The emergence of pseudo-magnetic fields and their corresponding Landau levels is more general than the low energy description of Weyl semimetals~\cite{Rachel2016}, yet it is in these systems where they take the very familiar form Eq.~\eqref{eq:B_5}.

The constant $b_0$, setting the shift of the Weyl nodes from zero energy,
plays the role of the scalar potential, motivating the definition of the four vector  (see Fig.~\ref{fig:types})
\begin{equation}
b^\mu=(b_0,\textbf{b}),
\end{equation}
where $\mu=0,x,y,z$ and the pseudo-electric field 
\begin{equation}
\label{eq:E_5}
    \textbf{E}_5=\partial_t{\textbf{b}}-\nabla b_0. 
\end{equation}

It is important to stress that the axial vector and scalar potentials in $b^{\mu}$ and the electromagnetic vector potentials are different~\cite{Landsteiner2014}. 
While the latter are gauge dependent quantities and hence not observable, the former are band structure parameters, ultimately quantum expectation values, and thus observable and gauge-invariant. 
The variation of $b^{\mu}$ in space is constrained to the region occupied by the crystal, implying that over the whole sample the average of the pseudo magnetic field is zero.

Chiral vector couplings emerge in elementary particle physics.
In the theory of electroweak interactions~\cite{Bertlmann2000,Peskin1995}, concerned with the interactions of neutrinos, for example, left and right chiralities couple differently to the gauge fields, and thus it is considered a chiral theory. However, the coupling is not to vector fields as above ($U(1)$ fields), but rather to upgraded to non-commuting matrices (non-Abelian fields). The simple vectors $\mathbf{B}_5$ and $\mathbf{E}_5$ have now found a first realization in condensed matter, and hence are the focus of this review. 

\section{\label{sec:Physorg}Physical origin of pseudo-fields}

\begin{table}
\begin{center}
\begin{tabular}{c|c|c|c|}
&$\mathcal{T}$&$\mathcal{I}$&  physical meaning\\
\hline
% $m$ & yes & yes & yes & Band gap when $b_\mu =0$\\
$\mathbf{b}$ & \xmark & \cmark &   magnetization \\
$b_0$& \cmark & \xmark & axial chemical potential\\
$\mathbf{B}_5$& \xmark & \cmark & pseudo-magnetic field\\
$\mathbf{E}_5$& \cmark & \cmark & pseudo-electric field
\end{tabular}
\end{center}
\caption{Inversion ($\mathcal{I}$) and time-reversal ($\mathcal{T}$) symmetry properties and a possible physical meaning of the parameters resulting in momentum ($\mathbf{b}$) and energy ($b_0$) separation of a Weyl semimetal. The symmetries of the corresponding pseudo-magnetic and electric fields are listed below.}
\label{table1}
\end{table}%

The discrete symmetries of Eq.~\eqref{eq:WeylH} help us reveal the physical requirements for the pseudo-fields to arise as a consequence of the modification of the Weyl node position. 
On the one hand, the coupling $\mathbf{b}\cdot\sigma$ is the familiar Zeeman coupling. Hence $\mathbf{b}$ can be often taken to be a magnetization breaking time reversal symmetry.
On the other hand, as far as discrete symmetries are concerned, $b_0$ enters as a chemical potential, and thus breaks space inversion symmetry. 
Then, using Eqs.~\eqref{eq:B_5} and \eqref{eq:E_5} we can reveal how the pseudo-fields transform under timer-reversal and inversion symmetries, summarized in Table~\ref{table1}. 
Similar to the magnetic field, $\mathbf{B}_5$ breaks time reversal symmetry while preserving inversion. The pseudo-electric field $\mathbf{E}_5$ preserves both symmetries, unlike the real electric field $\mathbf{E}$, which breaks inversion.

Two copies of the Weyl Hamiltonian Eq.~\eqref{eq:WeylH} with opposite signs of $\mathbf{b}$ can restore time-reversal symmetry, at the expense of inversion. 
An important situation where a doubling occurs is in Dirac semimetals where the two chiralities occur the same point in energy-momentum space, a degeneracy protected by an additional symmetry, for example, spin-rotation. Such symmetry can be present in a homogeneous material, but be  broken in a strained system due to emergent inversion symmetry breaking and spin-orbit interaction~\cite{Wang2012dirac,Wang2013three}. This symmetry breaking opens a gap near Dirac point. Although this masks the low-energy effects of the pseudofields, further away from the Dirac point the physics we described below can still apply. 
% %

Strictly speaking one can define a single unique vector $b^\mu$  only for the simplest Weyl semimetal with two Weyl nodes, as then it is the difference between the positions of the two nodes. For the generic multi-Weyl case with $n$ Weyl nodes, currently there is no unique way to define pseudo-fields.  
The presence of symmetries such as time reversal or crystalline symmetries will impose constraints linking pseudo-fields between pair of nodes. For example, in the case of a Dirac semimetal with two Dirac nodes composed of pairs of Weyl nodes connected by time reversal, the separation between Dirac nodes defines a single pseudo-field.

Irrespective of time-reversal and bearing in mind the multi-node subtleties above, strain couples directly to the Weyl node position, generating pseudo-fields. 
This coupling happens since strain modifies or introduces new overlap integral between atoms in a material, generically modifying the band structure (see Box 1)~\cite{bir1974symmetry,Shapourian2015,Cortijo2015}. At the tight-binding level, smooth strain profiles are introduced into a model via modifications of the hopping parameters determined by the components of the symmetrized strain tensor $u_{ij}=\frac{1}{2}(\partial_iu_j+\partial_ju_i)$ with $\textbf{u}$ being the local displacement vector of the strained lattice. A clever choice of $u_{ij}$ results in a displacement of the Weyl points in momentum space which translates into axial vector fields (see Box 1). Ref.~\cite{Cortijo2015} demonstrated that this is the case at the field theory level showing that strain parameters translate to gauge fields in the effective active action of the Weyl fermions. 

Inhomogeneous strain can have different physical origins. For example, in systems grown on a substrate, a lattice mismatch at the interface can create strain that gradually relaxes away from the interface~\cite{Grushin2016}. 
A different choice is to engineer wires twisted about their main axis, as proposed in Ref.~\cite{Pikulin2016}, leading to a constant bulk $\mathbf{B}_5$, as does bending a thin film to create a constant curvature (see ~\cite{Liu2017,Pikulin2018} and Fig.~\ref{fig:pumping}).
Other band structure parameters, such as the Fermi velocity, will also be modified as a result of such deformations (see ~\cite{Jiang2015,Yang2015,Arjona2017} and Box 1). 
Alternatively, a strain field is also induced by lattice defects. A screw-dislocation is a typical example where the strain gradient gradually relaxes away from it, which in turn can be reformulated as a pseudo-magnetic field parallel to the screw axis~\cite{Parrikar2014,Sumiyoshi2016,Chernodub2017}.

A change in position of the Weyl nodes in the Brillouin zone can also be driven by inhomogeneous time reversal breaking perturbations, through a space dependent Zeeman coupling $\mathbf{b}\cdot \sigma$. Static scenarios include magnetic domains~\cite{yamaji2014metallic,ma2015mobile}, magnetic vortices or skyrmions~\cite{Liu2013},
% , where the Weyl nodes separation in spatially modified because of a space dependent Zeeman coupling 
or exchange interactions~\cite{Shekhar2018,Cano2017}. 

A time dependent nodal separation, leading to the creation of a pseudo-electric field, can follow from dynamical strain. One of the first  examples that linked time dependent strain to the motion of the the Weyl points was presented in Ref.~\cite{Cortijo2016}, demonstrating that a rapid compression of a Weyl semimetal can result in an non-equilibrium chiral charge distribution due to a sudden shift of the node position in energy. Acoustic waves (phonons) are an alternative realistic source of time dependent perturbations that lead to pseudoelectric fields~\cite{Song2016,Pikulin2016,Rinkel2017}.

The most common and natural place to encounter pseudo-fields in Weyl semimetal is at their boundary, the place where the Weyl node separation must change abruptly from being finite to zero via the pair-wise annihilation of the Weyl nodes. This immediately implies, for surfaces onto which the Weyl node separation vector has a finite projection, a resulting strong pseudo-magnetic field confined to the surface. It is therefore not surprising that there is an intimate connection between the surface states of topological semimetals, namely the Fermi-arcs, and pseudo-fields~\cite{Grushin2016}. In the following Sec.~\ref{sec:B5}, we review the consequences, usages and generalizations of reinterpreting Fermi-arcs as the lowest Landau level of the strong axial field confined at the surface.

\section{\label{sec:B5}Pseudo-magnetic fields and pseudo-Landau levels}

\begin{figure*}
    \centering
    \includegraphics[width=0.8\linewidth]{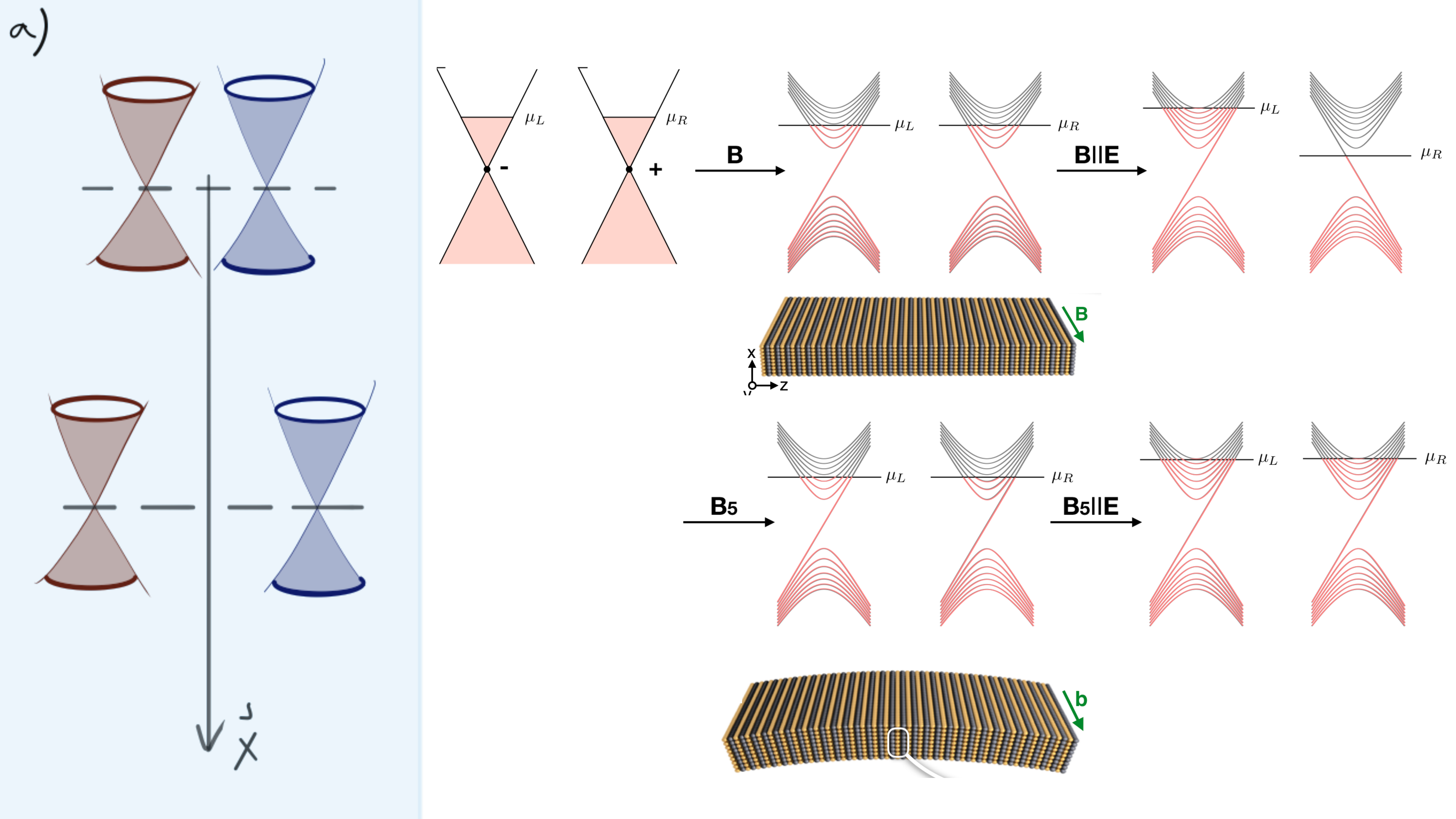}
    \caption{\textbf{Real and pseudo-magnetic fields}. The generation of a pseudo-magnetic field is achieved by varying the Weyl cone separation in space. The conical dispersion that defines the Weyl nodes splits into Landau (top panels) and pseudo-Landau levels (bottom panels) when the Weyl semimetal  a magnetic field $\mathbf{B}$ or a pseudo-magnetic field $\mathbf{B}_5$ are applied, respectively. The lowest Landau levels from the two cones are co-propagating for the real field and counter-propagating for the pseudo-field. Applying an external magnetic field creates charge pumping: combined with $\textbf{B}$ it is the standard inter node that results in chirality imbalance. Combined with $\textbf{B}_5$ it does not result in chirality imbalance, but charge is pulled from regions of opposite $\textbf{B}_5$ via the anomalous Hall effect.  }
    \label{fig:pumping}
\end{figure*}

Both the external magnetic field ($\mathbf{B}$) and a pseudo-magnetic field ($\mathbf{B}_5$) break the energy bands of the Weyl semimetal into Landau levels (see Fig.~\ref{fig:pumping}).
In the case of the external magnetic field, minimal substitution of an external vector potential $\textbf{A}$ in Eq.~\eqref{eq:WeylH} corresponding to a finite and constant magnetic field leads to a Landau level typical of relativistic fermions. The Landau energies depend on two quantum numbers: the Landau level index $n$ and the momentum along the direction of the magnetic field $k$. The energies for states with $n\neq 0$ are $E_n(k)=\mathrm{sign}(n)\hbar v_F\sqrt{2ne|\mathbf{B}|+k^2}$. The peculiarity of the nodal point is captured by the lowest (or zeroth) Landau level, which is linear in momentum and independent of the magnitude of the magnetic field. Its dispersion, $E_0=\chi \mathrm{sign}(|\mathbf{B}|) v_F k $ for a chirality $\chi$, shows that their direction of propagation is set by both by the direction of the magnetic field as well as the chirality of the node. For nodes of opposite chirality, they are \textit{counter-propagating}, namely they have an opposite group velocity with respect to one another. 
The band-structure is sketched in Fig.~\ref{fig:pumping}. 

There are two main effects associated with the existence of the chiral lowest Landau levels. 
One is the chiral anomaly, which we will consider in detail in Sec.~\ref{sec:anomaly}, and the second one is the chiral magnetic effect. 
The chiral magnetic effect is a current $\mathbf{J}$ that flows parallel to an applied magnetic field; $\mathbf{J}=\alpha \mathbf{B}$~\cite{Kharzeev2011,Son2012}. 
For a given Weyl node, the lowest Landau level of chirality $\chi$ contributes the number of filled states $\mu_\chi$, times the Landau level degeneracy proportional to $\mathbf{B}$, resulting in $\mathbf{J}_\chi=\chi(\mu_\chi/2\pi^2)\mathbf{B}$. 
The $n\neq0$ Landau levels do not contribute to the chiral magnetic effect since there are equal number of left and right movers at the Fermi energy.
Hence, a single Weyl node generates a current parallel to a magnetic field, the chiral magnetic effect, determined by its zeroth Landau level. 
In a lattice system such a current, however, cannot exist in equilibrium 
since the total chirality must sum up to zero, as shown by Nielsen and Ninomiya~\cite{Nielsen1981,Nielsen1981B}.
Counter propagating Landau levels always come in pairs with equal chemical potential, resulting in $\sum_{\chi}\mathbf{j}_\chi=0$.
The vanishing of this current in equilibrium is expected on general grounds, since it would lead to a charge build up that will eventually screen the current~\cite{Vazifeh2013,Landsteiner2014,Zubkov2016}.

Can a pseudo-magnetic field result in a chiral magnetic effect?
As follows from Eq.~\eqref{eq:WeylH}, Weyl fermions of opposite chirality $\chi$ experience 
the same magnitude of the field, but with an opposite sign. 
This has an immediate and striking consequence for the Landau levels described above: while the $n\neq0$ remain unaltered, the lowest landau level corresponding to each chirality are now \textit{co-propagating} (see Fig.~\ref{fig:pumping}). Their group velocity is fixed by the direction of the motion of the node in real space, i.e by the sign of pseudo-field alone. 

The appearance of co-propagating lowest Landau levels has consequences for the chiral magnetic effect. Per our discussion above, the current per node is $\mathbf{J}_\chi=\chi(\mu_\chi/2\pi^2)(\chi\mathbf{B}_5)$~\cite{Zhou2013,Sumiyoshi2016,Grushin2016,Huang2017}, which does not vanish when summing over chiralities in equilibrium. This implies a pseudo chiral magnetic effect in equilibrium which, unlike the chiral magnetic effect, is allowed. It is allowed because $\mathbf{j}\propto \mathbf{B}_5$ is not a transport current, but a magnetization current. To see this, note that the vector $\mathbf{b}$ transforms physically as magnetization and Eq.~\eqref{eq:B_5} is its curl, in parallelism with the definition of magnetization current~\cite{Jackson1998}. Such currents compensate each other between regions in the system with opposite signs of $\mathbf{B}_5$ and integrate to zero. Thus, as opposed to the chiral magnetic effect, that must vanish in equilibrium as a result of compensation of modes in momentum space, the pseudo chiral magnetic effect sums up to zero when integrated over regions \textit{in real space}.

Pseudo-magnetic fields offer a perspective on the physical origin of the surface states of Weyl semimetals, the Fermi-arcs. The surface represents the place where Weyl nodes annihilate in pairs, and $b^\mu$ vanishes, rapidly changing the nodal separation over a short distance. It is this observation that was used in Ref.~\cite{Grushin2016} to propose, supported by lattice calculations, that the Fermi arc is secretly the lowest Landau level of a strong $\textbf{B}_5$ confined to and lying in the surface plane, an idea that was later developed further to include surface states in tilted and over tilted Weyl nodes~\cite{Tchoumakov2017}. It is a generalization of the view that when a band parameter varies in space smoothly, such as the mass, new levels are allowed to occur, that we can now reinterpret as effective Landau levels~\cite{Rachel2016}, as in topological insulators with smooth interfaces~\cite{Tchoumakov2017,Inhofer2017}.

Viewing Fermi arcs as the lowest, chiral Landau level associated to $\mathbf{B}_5$ eliminates the artificial dichotomy between bulk and surface. We are allowed to regard a Weyl semimetal as a path traced by a pair of nodes. They are created from the vacuum at one surface and annihilated at the opposite surface. The variation of their relative distance defines locally a $\mathbf{B}_5$ and the pseudo chiral magnetic effect determined by the direction of the Landau level’s group velocity. Any bulk uniform pseudo-chiral magnetic effect is compensated by an imbalance of Fermi arc lengths~\cite{Grushin2016}.

The similarities and differences between pseudo Landau levels and ordinary Landau levels are explicit when depicting the Fermi surface in the presence of both types of fields in a finite sample~\cite{Bulmash2016,Ominato2016} (see Box 2). 
In the quantum limit the Fermi surface of a Weyl semimetal in a magnetic field $\mathbf{B}$ is a parallelogram in the two dimensional momentum space perpendicular to the magnetic field.
Two sides of the parallelogram represent the position momentum locked manifold of states of the bulk zeroth Landau level, while the additional two sides are the Fermi arcs, or equivalently, the zeroth pseudo Landau level resulting from the finite $\textbf{B}_5$ at the surface. More general profiles of $\textbf{B}_5$ deform the Fermi surface which can take, among others, a peculiar bow-tie shape~\cite{Behrends2018}, which we discuss in Box 2. 

Finally, we point out that such combination of an external magnetic field with an intrinsically emerging $\textbf{B}_5$ can yield interesting twists over known effects. 
In the presence of a bulk $\textbf{B}_5$, the particles experience the total field $\textbf{B}_{\chi}=\textbf{B}+\chi\textbf{B}_5$, giving rise to modified bulk trajectories that can affect quantum oscillations~\cite{potter2014quantum,zhang2016quantum,Pikulin2018}. 
We note as well that being an intrinsically emergent band structure property, $\mathbf{B}_5$ can couple to chargeless degrees of freedom, unlike an external magnetic field.
This allows Landau levels in the spectrum of Bougoliubov quasiparticles~\cite{Nica2018} and pressure or photonic waves in Weyl metamaterials~\cite{Peri2018}.

\section{\label{sec:E5}Pseudo-electric fields}

\begin{figure*}
    % \centering
    \includegraphics[width=0.8\linewidth]{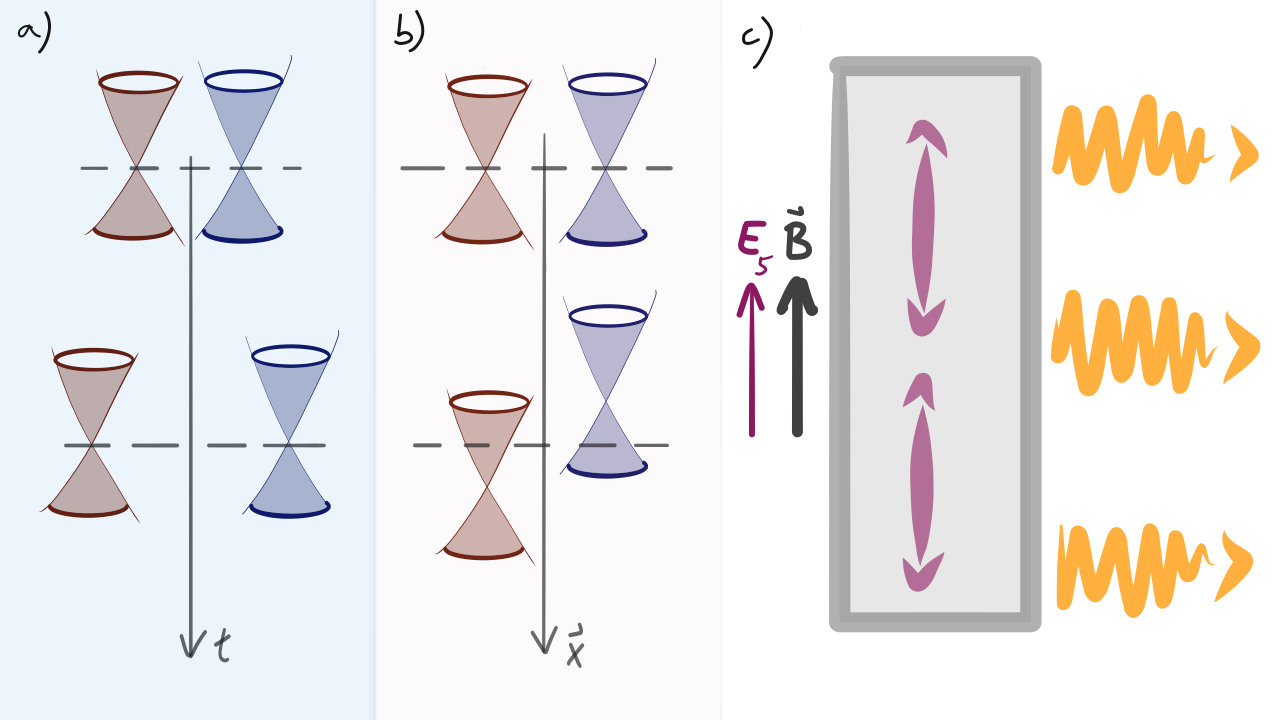}
    \caption{\textbf{Pseudo-electric field and a physical consequence}. (a) and (b) illustrate the two inequivalent ways to generate a pseudo-electric field. (a): The relative position of the Weyl nodes in energy changes as a function of time. (b): The relative position of the Weyl nodes in momentum space changes as a function of position. (c) shows an example of the signature of the pseudo-electric field: a Weyl semimetal in an external electric field in the presence of a longitudinal sound wave propagating through the sample. The chiral magnetic effect leads to charge redistribution in the sample and emanating electromagnetic radiation~\cite{Pikulin2016}.}
    \label{fig:E5}
\end{figure*}

Similar to the pseudo-magnetic field discussed above, a pseudo-electric field can be extracted from Eq.~\eqref{eq:E_5}. The two terms in this equation set two different ways to generate $\mathbf{E}_5$. The first is by a spatial variation of the separation of Weyl nodes in energy, the second is through a time variation of the separation of Weyl nodes in momentum (see Fig.~\ref{fig:E5}). 

As we mentioned above, the two ways of generating pseudo-electric fields are not equivalent.
Unlike the gauge potential and vector potential, both $\mathbf{b}$ and $b_0$ are physically observable and they are not transformed into each other under a gauge transformation.
Physically, a system subjected to a spatial variation of $b_0$ remains in equilibrium while a time variation of $\mathbf{b}$ drives the system out of equilibrium.
Quenching $b_0$ in time, suddenly from zero to a finite value, will create a chiral chemical potential imbalance between the two nodes~\cite{Cortijo2016} for times shorter than the equilibration time, the inter-node scattering time. 
During this short time (of the order of a picosecond) a chiral chemical potential imbalance ($\mu_+\neq \mu_{-}$) can exist.
Following the discussion in the previous section, a current parallel to an applied external magnetic field emerges: a chiral magnetic effect induced by the quench.

When pseudo-electric and pseudo-magnetic fields coexist, Ref.~\cite{Arjona2017} predicted a strain induced pseudo-Landau level collapse. It generalizes the Landau level collapse that occurs when perpendicular electric and magnetic fields are present, and satisfy $|\mathbf{E}|=v_F |\mathbf{B}|$, such that a boost to reference frame where $\mathbf{E}=0$ yields $\mathbf{B}=0$. At this critical field, the energies of consecutive Landau levels coincide.
A similar effect was predicted for a pseudo magnetic-field~\cite{Arjona2017}, where a collapse of its pseudo-Landau levels occurs due to the presence of a pseudo-electric field, at an analogous critical value $|\mathbf{E}_5|=v_F |\mathbf{B}_5|$. This prediction, valid in the low energy limit, is consistent with an independent boost in the two Weyl nodes and further solidifies the analogy between the pseudo- and real fields.

\section{\label{sec:semiclassical}Pseudo-fields in the semi-classical approximation}
The semi-classical formalism provides a source of classical intuition into the quantum reality, and has become a popular approach in the context of topological metals. 
The brilliant realization by Berry that the Bloch wave-functions can acquire a topological phase winding over the Brillouin zone, upgrades the semi-classical equations of motion to include an anomalous velocity term~\cite{Sundaram1999,karplus1954hall,nagaosa2010anomalous,Xiao2010}.
The anomalous velocity depends on the Berry curvature, $\boldsymbol{\Omega}$, defined from derivatives of the wave-function, and enters the semiclassical equations of motion as a magnetic field in reciprocal space \cite{Xiao2010}.  

From the previous two sections it follows that a natural starting point to account for the effect of pseudo-fields is to consider two independent Fermi surfaces close to Weyl nodes that feel different effective electromagnetic fields depending on their chirality $\chi=\pm$. Then, following Refs.~\cite{Pikulin2016,Grushin2016}, one can write an equation for each node subject to the effective fields $\mathbf{B}_{\chi}=\mathbf{B}+\chi\mathbf{B}_5$ and $\mathbf{E}_{\chi}=e\mathbf{E}+\chi\mathbf{E}_5$
\begin{subequations}
\label{eq:semiclassics}
\begin{eqnarray}
\label{eq:semiclassics1}
\dot{\mathbf{r}}_\chi &=& \dfrac{\partial \epsilon_\chi}{\partial \mathbf{k}}  + \mathbf{\Omega}_\chi\times \dot{\mathbf{k}}_\chi,\\
\label{eq:semiclassics2}
\dot{\mathbf{k}}_\chi &=&  \mathbf{E}_\chi + \dot{\mathbf{r}}_\chi\times \mathbf{B}_\chi.
\end{eqnarray}
\end{subequations}
where $\mathbf{r}_\chi$ and $\mathbf{k}_\chi$ represent the coordinate and momentum around which a semiclassical wavepacket is centered. 

Although inserting $\mathbf{E}_\chi$ and $\mathbf{B}_\chi$ this way into the equations of motion seems like a reasonable guess, one might worry that these equations do not immediately follow from semi-classical theory, and that the pseudo-fields have been added heuristically based on the physical intuition developed in the previous sections. 
When perturbations are inhomogenous, such as in the case of strain, more sophisticated and complete semiclassical equations are needed~\cite{Sundaram1999}. 
The usual momentum space Berry curvature is then part of a more general tensor, $\Omega^{ab}$, where $a,b$ indicate that the two derivatives of the wave-function that define the Berry curvature are to be taken with respect to momentum $\mathbf{k}$, position $\mathbf{r}$ and time $t$.
From these general semiclassical equations, and provided one assumes the momentum of the wave-packet is close to a Weyl node, a change of coordinate frame recovers Eqs.~\eqref{eq:semiclassics}~\cite{Roy2018}, reconciling them with standard semiclassical theory~\cite{Sundaram1999}.
In this frame, the wave-packet motion is measured from the nodes' position in momentum space, turning spatio-temporal variations of the Weyl node separation $b_\mu$ into the pseudo-fields.

From the semiclassical Eqs.~\eqref{eq:semiclassics} we can predict simple physical effects for wave-packet motion in the presence of pseudo-fields~\cite{Roy2018}.
As a first example, the last term in Eq.~\eqref{eq:semiclassics2} implies that $\mathbf{B}_5$ imprints cyclotron orbits with opposing senses for each chirality, indicative of the fact that they feel an opposite effective magnetic field. 
Interestingly, this has implications for a propagating wave-packet that crosses a region where the node separation changes, such as the interface between two Weyl semimetals.
As it transverses the boundary, it will experience the interface pseudo-magnetic field $\mathbf{B}_5$ and will be deflected in a direction that depends on the sign of $\mathbf{B}_5$ at the boundary, or equivalently, on the dispersion of the Fermi-arc at the interface.~\cite{Roy2018}.

As a second example, the combination of the electric field term of Eq.~\eqref{eq:semiclassics2} and the anomalous velocity term in Eq.~\eqref{eq:semiclassics1} implies that $\mathbf{E}_5$ can create a pseudo-Hall effect, proportional to $\boldsymbol{\Omega}\times \mathbf{E}_5$, since the product of the Berry curvature and velocity due to $\mathbf{E}_5$ is chirality independent, and thus does not vanish once we sum over nodes~\cite{Roy2018,Grushin2016,Pikulin2016}.
These predictions can be checked numerically in lattice models and are well suited for testing in cold-atomic~\cite{Roy2018}, acoustic or photonic realizations of Weyl semimetals (see section \ref{sec:experiments}).

\section{\label{sec:anomaly} The chiral anomaly with pseudo-fields}

One of the peculiar features of Weyl fermions is the emergence of the chiral anomaly in the presence of external electromagnetic fields~\cite{Bertlmann2000}.
Pseudo-electromagnetic fields enter the chiral anomaly, yet their intrinsic origin, as opposed to the external electromagnetic fields, severely modifies its phenomenology.

The current understanding of the chiral anomaly including intrinsic pseudo-fields is a blend between semiclassical approximations, quantum field theory and tight binding calculations. Semi-classical approaches have the benefit of being physically intuitive  yet it is not immediately obvious that they are powerful enough to describe anomalies. Pioneering work by Son and Spivak~\cite{Son2013} showed that combining the semi-classical equations of motions for a wave-packet of electrons with the Boltzmann equation, can recover the anomaly equations in the absence of $\mathbf{E}_5$ and $\mathbf{B}_5$.
As a spin-off, they obtained a positive and increasing longitudinal magneto-conductivity as a function of magnetic field strength, which is currently taken as the fingerprint of the chiral anomaly~\cite{Son2013}.
Is it possible then to recover the anomaly equations with fields and pseudofields~\cite{Bertlmann2000} within the semiclassical approximation, and they lead to increased conductivity? 

The efforts to derive the chiral anomaly in a semi-classical approximation, generalizing Ref.~\cite{Son2013} to include pseudo electromagnetic fields reveals a few subtleties of this calculation.
Combining Eqs.~\eqref{eq:semiclassics} with the kinetic equation recovers the Fermi surface contribution to the chiral anomaly, referred to in high energy physics as the covariant anomaly~\cite{Liu2013,Landsteiner2014,Landsteiner2016}.
The covariant anomaly relates the conservation of the axial current $J^{\mu}_5=J^{\mu}_L-J^{\mu}_R$, measuring the difference between currents of left and right chiralities, and the total current, $J^{\mu}=J^{\mu}_L+J^{\mu}_R$, to electromagnetic fields and pseudo-fields by the following equations
\begin{eqnarray}
\label{eq:chiralanomalymain1}
    \partial_\mu J^\mu_5=\frac{1}{2\pi^2}(\mathbf{E}\cdot \mathbf{B}
    +\mathbf{E}_5\cdot \mathbf{B}_5),\\
    \label{eq:chiralanomalymain2}
   \partial_\mu J^\mu=\frac{1}{2\pi^2}(\mathbf{E}\cdot \mathbf{B}_5
   +\mathbf{E}_5\cdot \mathbf{B}).
\end{eqnarray}
Strikingly, Eq.~\ref{eq:chiralanomalymain2} breaks charge conservation in the presence of pseudo-fields.
Since charge must be conserved, it follows that additional currents must exist in the system to restore it, that are not captured by the semiclassical approximation.
Another drawback of the semiclassical, even without pseudo-fields, is that it misses the Hall current proportional to the Weyl node separation, a known response of Weyl systems~\cite{Burkov2011,Burkov2014,Armitage2018}.
This, in retrospect, is expected since the semiclassical approximation accounts for the single band Fermi surface properties of wave-packets, and the Hall conductivity carries information of all filled states.

To solve these problems, the authors of Ref.~\cite{Gorbar2017c} fixed the semi-classical theory in the presence of pseudo-fields by adding extra currents to $J^\mu_5$ and $J^{\mu}$, known in the high energy literature as Bardeen Polynomials~\cite{Bardeen1984,Landsteiner2016}.
They account for the loss of charge and allow to recover the current conserving version of the anomaly, known as consistent anomaly.
In the resulting \textit{consistent} chiral kinetic theory, the semiclassical picture is complemented by currents that are beyond the cut-off of the low energy theory, in such a way that charge is conserved. 
One of these currents is precisely that Hall current proportional to the Weyl node separation, recovering what is expected from lattice calculations~\cite{Burkov2011}.
In Box 3 we present a summary of the quantum field theory viewpoint on consistent and covariant anomalies and give some details of the connection of Bardeen polynomials to lattice models.

Eqs.~\eqref{eq:chiralanomalymain1} and \eqref{eq:chiralanomalymain2} and their consistent versions (see Box 3), open the exploration of a number of important physical consequences tied to the interplay between real and pseudo-fields. 
The anomaly term generated purely by external fields, $\mathbf{E}\cdot\mathbf{B}$ in Eq~\eqref{eq:chiralanomalymain1}, causes an inter-valley charge pumping that translates to an imbalance of chemical potential at nodes of opposite chirality (see~\ref{fig:pumping}).
In contrast, terms mixing external and pseudo fields in Eq.~\eqref{eq:chiralanomalymain2} can pump charge between different regions in real space. 
Figure \ref{fig:pumping} sketches an example for such a process, resulting from a combination of a bulk $\textbf{B}_5$ and an external electric field. 
In this case, the electric field adds (or removes) charge near the two nodes by an equal amount.
To preserve charge conservation, the charges must be drawn from another region where the sign of $\textbf{B}_5$ is reversed, possibly at the boundary. 
A similar charge redistribution is expected from applying an external $\text{B}$ not orthogonal to an intrinsic $\mathbf{E}_5$, yet it is important to acknowledge the two in-equivalent ways of generating $\mathbf{E}_5$ discussed in Sec.~\ref{sec:E5}.
Since a static gradient of $b_0$ cannot cause charge dynamics, the statement of Eq.~\eqref{eq:chiralanomalymain2} is then trivial: the Fermi energy does not depend on position for a static system. 
However, if $\mathbf{E}_5$ is generated by a time-dependent perturbation, Eq.~\eqref{eq:chiralanomalymain1} and \eqref{eq:chiralanomalymain2} describe dynamical charge redistribution.

Phonons, which are time dependent lattice distortions, are a natural way to create $\mathbf{E}_5$ dynamically.
They can couple to a magnetic field through the $\mathbf{E}_5\cdot \mathbf{B}$ term in Eq.~\eqref{eq:chiralanomalymain2} that modifies their dispersion at low frequency \cite{Song2016,Spivak2016, Pikulin2016,Rinkel2017}. 
From the functional form $\mathbf{E}_5\cdot \mathbf{B}$ it is apparent that such coupling is most pronounced when a longitudinal homogeneous acoustic wave is propagating along the direction of an external magnetic field applied parallel to the Weyl node separation. 
In this case, the acoustic wave causes charge redistribution between the bulk and the boundary of the material, thus resulting in emitted electromagnetic radiation with the wavelength of the acoustic wave \cite{Pikulin2016}, see also Fig~\ref{fig:E5}.

The coupling of real and pseudo electromagnetic fields predicts the existence of chiral magnetic waves~\cite{Kharzeev2011, Gorbar2017}. 
They are a collective excitation of Weyl semimetal in presence of an external magnetic field, enabled by the combination of the two chiral anomaly equations Eqs.~\eqref{eq:chiralanomalymain1} and \eqref{eq:chiralanomalymain2} understood as follows.
From Eq.~\eqref{eq:chiralanomalymain1}, $\mathbf{E}\cdot\mathbf{B}$ causes a chiral chemical potential difference, leading to a finite chiral current $\mathbf{J}_5$. 
The chiral current creates a time dependent gradient of chiral chemical potential $b_{0}$, leading to $\mathbf{E}_5$. 
Consequently, $\mathbf{E}_5\cdot \mathbf{B}$ in Eq.~\eqref{eq:chiralanomalymain2} creates a current $\mathbf{J}$, which finally creates a gradient of electric potential and thus $\mathbf{E}$.

In transport the most notable experimental consequence is a strain induced enhancement of the longitudinal conductivity as a function of the pseudo-magnetic field strength~\cite{Pikulin2016,Grushin2016}.
In analogy to \cite{Son2013}, this result was obtained within the semiclassical approximation by combining Eq.~\eqref{eq:chiralanomalymain1} and the Boltzmann transport equation, and can be written schematically as $\sigma\propto |\mathbf{B}_5|^3$ for uncorrelated short-range disorder and chemical potential away from the Weyl node~\cite{Pikulin2016}.
This prediction remains to be observed in experiment.

An open problem with the semiclassical approach is that
it often assumes that the chiral chemical potential imbalance resulting from the chiral anomaly is compensated by an inter-node scattering mechanism.
In the presence of pseudo-fields the description of such scattering time is challenging~\cite{Gorbar2017c} since it seems to involve regions with opposite signs of the pseudo-magnetic field~\cite{Grushin2016,Pikulin2016,Behrends2018, Landsteiner2018}.
A complete scattering theory in the presence of pseudo-fields that can clarify this issue is currently absent.

To conclude this section, it is interesting to point out that there are other sources of anomalies related to strain.
First, the presence of dislocations also leads to inhomogeneous strain gradients~\cite{Zubkov2015,Yizhi2016,Chernodub2017,Huang2018,Ferreiros2018,SotoGarrido2018}. 
Using a variety of complementary methods, most notably quantum field theory, it is possible to show that a density of screw dislocations leads to torsional fields~\cite{Ferreiros2018,Huang2018}, which contribute to the chiral anomaly and have observable consequences.
A recent example is the activation of a mixed axial-torsional anomaly by sound waves propagating through a Weyl semimetal 
with a strain induced spatial variation of the tilt of the Weyl cone dispersion~\cite{Ferreiros2018}.
Within the semi-classical approximation treating the effect of dislocations on the chiral anomaly requires extending the commonly used approach of Ref.~\cite{Sundaram1999} as discussed in Ref.~\cite{Huang2018}.
Second, the combination of thermal gradients and pseudo-fields is sensitive to a more exotic contribution to the chiral anomaly, the mixed gravitational anomaly~\cite{Landsteiner2016,Gooth2017,Stone2018,Schindler2018}.
So far, this type of anomalies have only been probed experimentally applying real magnetic fields~\cite{Gooth2017,Stone2018,Schindler2018}.
However, they can be also probed using pseudo-fields. 
For example a Weyl semimetal will rotate under heating or cooling through the pseudo chiral magnetic effect~\cite{Chernodub2014}.
The rotation is caused by the existence of a term that couples angular momentum, $\mathbf{B}_5$ and a temperature gradient and proportional to the coefficient of the mixed gravitational anomaly~\cite{Chernodub2014,Landsteiner2016}.

\section{\label{sec:experiments}Experimental realizations and probes}
\begin{figure*}
    \centering
    \includegraphics[width=0.8\linewidth]{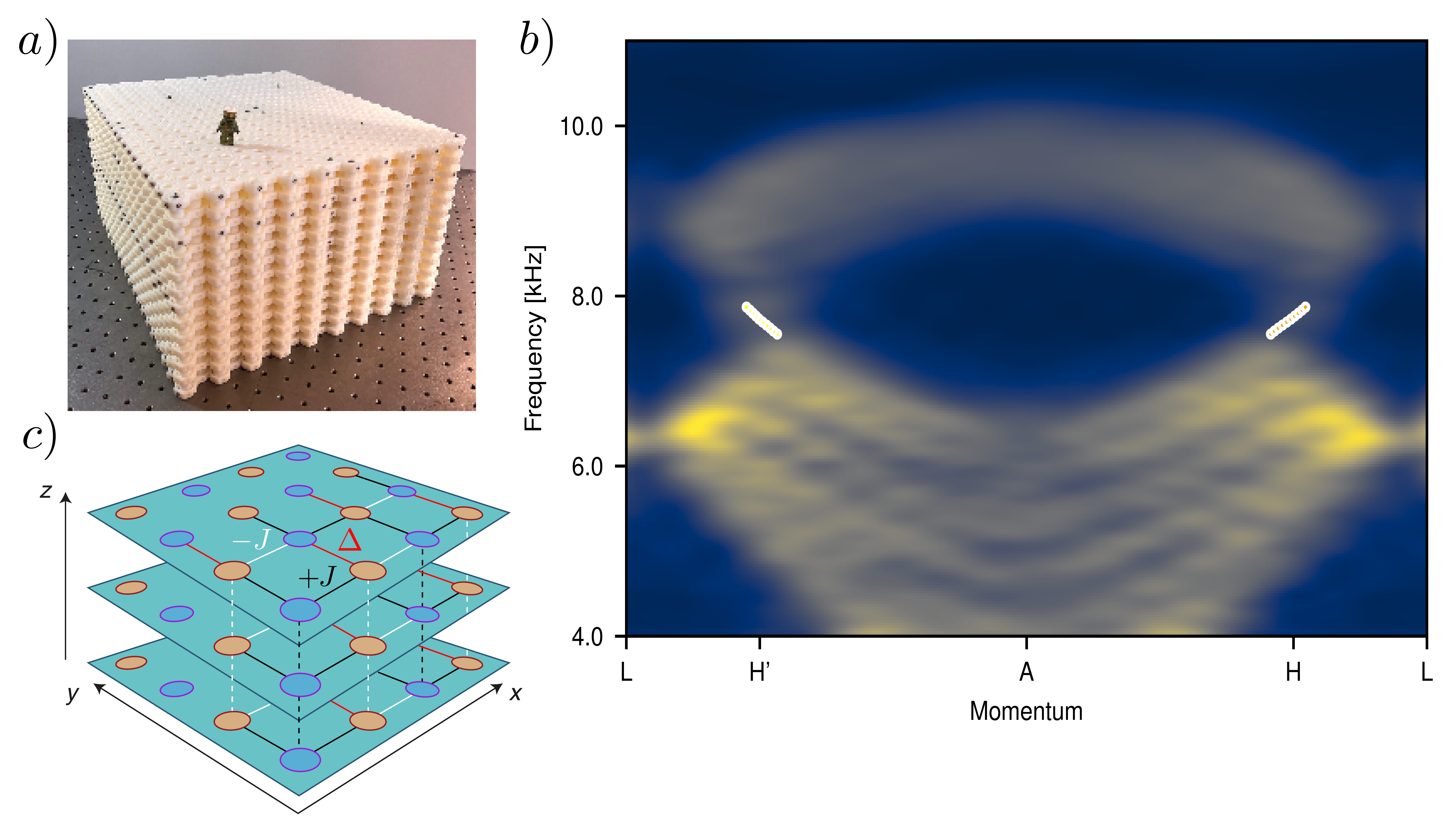}
    \caption{\textbf{Experimental realizations of pseudo-fields}. (a) The inhomogeneous acoustic Weyl material realizing a constant bulk pseudo-magnetic field used in Ref.~\cite{Peri2018}. The structure is rigid and made of stacked honeycomb lattices with propagating pressure waves. The size of the holes in the structure are varied in real space. This ammounts to a sublattice degree of freedom that varies in space, creating a space dependent Weyl node position. (b) The acoustic wave dispersion spectrum shows chiral pseudo-Landau levels (white) which have the same sign for nodes of opposite chiralities. H and H' are points in the Brillouin zone that are related to one another by time reversal symmetry. The chiralities of the Weyl nodes located at H and H' are the same. The hole inhomogeneity results in a real space motion of the nodes in opposite directions. Hence they must feel opposite pseudo-magnetic fields, and their lowest Landau levels are counter propagating. (c) Proposal for the realization of gauge fields in cold atomic lattices~\cite{Roy2018} based on dimerized tunneling amplitudes ($J,\Delta$) in the xy plane. By modulating the optical potential of Ref.~\cite{Tarruell2012}, $\Delta$ can be varied in space and time generating  axial fields.}
    \label{fig:exp}
\end{figure*}

The physical realization of axial fields is a thriving frontier of the field.
The generality of the existing formulation of axial fields promises that physically different systems can realize analogous physics 
(c.f. section \ref{sec:Physorg}).
As of today, the first and only realization of an axial magnetic field in three dimensions was achieved in an acoustic analogue of a Weyl semimetal~\cite{Peri2018} (see Fig.~\ref{fig:exp}).
Acoustic Weyl semimetals are three dimensional printed structures with a periodic pattern of cavities and holes~\cite{Xiao2015,Abbaszadeh2017}.
For certain frequencies, pressure waves are forced to propagate with a linear frequency and momentum relationship that mimics a Weyl dispersion.
In a prominent step forward, the authors of Ref.~\cite{Peri2018} used the versatility of 3D printing to design spatially modulated structures.
This modulation effectively varies the node separation in space, realizing the model proposed in Ref.~\cite{Roy2018}, implementing a constant 
axial pseudo-magnetic fields in the bulk of the sample.
In this set-up, pressure waves at a certain window of frequencies propagate via zero pseudo-landau levels that disperse in the same direction for nodes of opposite chirality, a quintessential signature of an underlying pseudo-magnetic field.

Other synthetic platforms are plausible alternatives to observe pseudo-fields.
For instance, realistic modifications of current ultra-cold atomic lattices have been predicted to host both pseudo electric and magnetic fields, controllable externally~\cite{Roy2018}.
Photonic lattices, the platform where Weyl nodes were first observed~\cite{Lu2015}, may allow for spatial modulation of the coupling between different waveguides~\cite{Rechtsman2013} inducing an effective node separation as a function of a given space coordinate.

In the solid state perhaps the simplest existing consequence of a pseudo-magnetic field are the topological Fermi arcs, as extensively discussed above.
Interestingly, for the 2D topological surface states in topological insulators strain and its pseudo-magnetic field result in flat bands~\cite{Tang2014} and non-topological surface states have been recently observed and interpreted as higher pseudo Landau levels in topological insulator smooth interfaces~\cite{Tchoumakov2017,Inhofer2017}.
However, higher pseudo Landau levels have not been reported for Weyl semimetals, yet they are expected as well.

Richer interface phenomena can lead to a higher degree of control over pseudo-magnetic fields and broader phenomenology. 
Consider for instance sharp or smooth interfaces between topological metals and insulators, which generalize the way Fermi arcs occur.
The boundary between two Weyl semimetals of different Weyl node separations hosts smaller Fermi arcs, given by the difference of the two node separations, linked as well to a pseudo magnetic field~\cite{Grushin2016}.
Moreover, interfaces between insulators (topological or trivial) and semimetals can simultaneously host both Dirac surface states and pseudo-Landau levels~\cite{Grushin2015,Lau17,Juergens2017}.
Going one step further, the concatenation of several types of these interfaces, in the form of heterostructures, can result in even richer phenomenology~\cite{Yesilyurt2017,Hills2017,Weststrom2017}.
The interface of two distinct Weyl semimetals is expected to result in electronic Veselago lensing~\cite{Hills2017,Weststrom2017}, which can be traced back to an emergent space-time metric for the Weyl fermions due to underlying strain~\cite{Weststrom2017}. 
The exact connection between the metric formulation~\cite{Zubkov2015,Cortijo2016b,Volovik2016,Guan2017,Nissinen2017,Zubkov2018} and the pseudo-fields, as that existing for graphene~\cite{deJuan2012}, remains to be formally established.
This connection can aid to realize effective black-hole metrics~\cite{Volovik2016,Nissinen2017} and serve as design principles for metamaterials~\cite{Weststrom2017}.

It would be desirable to realize bulk pseudo-fields rather than interface phenomena, similar to the acoustic implementation of pseudo-gauge fields~\cite{Peri2018}.
First, strain has been proposed as a way of realizing Weyl semimetals, in strained HgTe~\cite{Ruan2016} and in group-V materials~\cite{Moynihan2017}.
Strain profiles in these materials are likely to be inhomogeneous, generating pseudo-electromagnetic fields in the bulk, albeit with a lesser degree of control over other strain effects (see Box 1).
In particular, a pseudo-electric field can over tilt the Weyl dispersion~\cite{Alisultanov2018} (known as a Type-II Weyl node~\cite{Soluyanov2015}) and induce a finite magnetization~\cite{Alisultanov2018b}.
Second, it is plausible that by etching crystals into different geometries using lithography, used recently to discover the peculiar quantum oscillations mediated by Fermi arcs~\cite{potter2014quantum,Moll2016,Zhang2018z}, can be used to strain samples in a controlled way . 

Bulk inhomogeneous strains can be engineered, for example, 
in wires twisted about their main axis, as proposed in Ref.~\cite{Pikulin2016}. 
In this geometry, and when both magnetic and pseudomagnetic fields are applied, left- and right-handed Weyl fermions can be focused at different spatial locations, resulting in a chirality filter~\cite{Gorbar2017b}.

As mentioned in Sec.~\ref{sec:Physorg}, screw dislocations traversing the bulk of a 3D material are also a source of strain, and are predicted to result in quasi one dimensional currents flowing along the dislocation line. 
Such currents are referred to as a torsional chiral magnetic effect in Ref.~\cite{Parrikar2014,Sumiyoshi2016,Chernodub2017}. 
and are an instance of the pseudo chiral magnetic effect:
a current parallel to the pseudo-magnetic field generated along the dislocation line due to its associated strain field.
A related current generation occurs in the vicinity of magnetic vortices, where the magnetization is space dependent, generating an axial field~\cite{Liu2013}.

Regarding transport probes, the enhancement of the electrical conductivity~\cite{Pikulin2016,Grushin2016} or signature quantum oscillations~\cite{Liu2017,Pikulin2018}
as well the other effects described in previous sections,  can act as measures of the presence of pseudo-magnetic fields.

Additionally, novel collective excitations can emerge in the presence of a bulk pseudo-magnetic field~\cite{Gorbar2017e,Gorbar2017d}.
For instance, low energy gapless charge excitations known as helicons, which typically require the presence of magnetic fields, can exist as long as $\mathbf{B}_5\neq0$~\cite{Gorbar2017d}.

Dirac/Weyl superconductors or other nodal superconducting systems are another intriguing platform to explore the effects of the pseudo-fields. The single-particle Hamiltonian of such a superconductor is identical to the corresponding semimetal, the peculiarity of the Bogoliubov-de Gennes description~\cite{Tinkham2004} of the system allows for chiral magnetic effect in equilibrium.
It manifests itself as a supercurrent~\cite{OBrien2017} emerging along the pseudofield in the Dirac/Weyl superconductor~\cite{Matsushita2018}. This is unlike the usual Meissner effect, where the supercurrent flows perpendicular to the external field to screen it\cite{Tinkham2004}. It has also been shown that pseudofields affect the superconducting pairing type in Dirac/Weyl semimetals~\cite{Gorbar2018}: the p-wave superconductivity with the total Cooper pair spin parallel to the pseudo-field direction is the most favorable pairing.

A final intriguing possibility is to make use of light to generate pseudo-fields. 
Circularly polarized light breaks time reversal symmetry, shifting the Weyl node position~\cite{Chan2016, Sie2019} and generating an additional Weyl node separation $\mathbf{b}_\mathrm{light}$ that would add to the pre-existing $b_\mu$.
Since the intensity of light can be inhomogeneous, this can lead to an inhomogeneous $b_\mu$ profile resulting in pseudo-magnetic field. Light can induce a time-dependent strain of the Weyl material, and shift the Weyl node positions in time, resulting in pseudo-electric fields. 
A recent experiment suggests that such shift can be performed~\cite{Sie2019}.

\section{\label{sec:outlook}Outlook}

The physics of pseudo-fields confronts us with remarkable future challenges.
One direction to explore is the realization of pseudo-field in different generalizations of the Weyl and Dirac semimetals. In particular, materials with double and triple Weyl nodes~\cite{Fang2012,Wieder2016}, which are squared or cubic versions of Hamiltonian~\eqref{eq:WeylH} protected by rotational symmetries, are expected to display pseudo-field related phenomena~\cite{Huang2017,Sukhachov2018} and have no high-energy analogue.
Additionally, multi-fold fermions~\cite{manes2012,Bradlyn2016,Chang2017,Tang2017,Bouhon2017}, generalizations of Weyl nodes where multiple bands meet protected by lattice symmetries, can also realize emergent relativistic quasiparticles and corresponding pseudo-fields such as those considered in high energy physics and beyond.
%
% %
An example is a position-dependent three-fold fermion, which can carry an SU(3) symmetry and enable axial color anomalies, resembling chromodynamic theory. More generally four-, six- and eight-fold fermions go beyond what is available in high energy and are still to be exploited.

Another frontier is to experimentally realize pseudo-fields in a wide variety of systems. The first signatures have been reported in the metamaterials~\cite{Peri2018} which allow to carefully manufacture the required pseudo-fields. However, those are bosonic materials, thus lacking Fermi surface. This limits the number of observable signatures of the pseudofields, in particular the measurement of anomalies. To connect these to experiments, it is necessary to observe pseudo-fields in an electronic material, beyond the existence of Fermi arcs.

Pseudo-electromagnetic fields emerge in graphene, Weyl and Dirac semimetals. However, they are not restricted to low energy relativistic systems~\cite{Rachel2016}; any momentum-space-dependent field in principle can be treated as a pseudofield. Thus, we expect that more systems can show effects of emergent pseudo-fields. Those may include regular Landau-Fermi liquids with a position-dependent deformation of their Fermi surfaces, but also other candidate topological materials, including inhomogeneous nodal line semimetals or topological insulators. The current status of pseudo-electromagnetic fields in Weyl semimetals, which we have reviewed, is indicative of a nascent and rich subject of research, where we expect important breakthroughs to emerge.

\section*{Acknowledgments}

We are grateful to J.~H.~Bardarson, J.~Behrends, A.~Chen, A.~Cortijo, Y.~Ferreiros, M.~Franz, S. Huber, M.~Kolodrubetz, K.~Landsteiner, T.~Liu, P.~Moll, V. Peri, D.~Pesin,  S.~Roy, A.~Stern, A.~Vishwanath, J.~W.~F.~Venderbos and M.~A.~H.~Vozmediano for discussions and related collaborations. We further thank S. Huber for contributing to figure \ref{fig:exp} and J. Berhends, V. Peri and K. Landsteiner for their critical reading of the manuscript.
R. I. is supported by the ISF (grant No. 1790/18). A. G. G. is supported by the ANR under the grant ANR-18-CE30-0001-01, the Marie Curie programme under EC Grant agreement No.~653846 and de EC project FET-OPEN SCHINES No.~829044.

\appendix

\section*{BOX 1: From strain to pseudo-fields}

Let us illustrate the transformation of the strain to pseudo-fields using an example of a toy lattice model having two Weyl points in the Brillouin zone:
\begin{eqnarray}\nonumber
H(\mathbf{k})&=&t_\perp \sigma_x \sin k_xa +t_\perp\sigma_y \sin{k_ya} \\
&+& t_\parallel \sigma_z (\cos k_z a -m).
\end{eqnarray}
This model supports two Weyl points at $\pm |\mathbf{b}|=\pm\arccos{m}$ for any $-1<m<1$. For the case of isotropic Weyl nodes ($t_\perp = t_\parallel\sqrt{1 - m^2}$) the Fermi velocity in  Eq. \eqref{eq:WeylH} is $v_F=t_\perp a = t_\parallel a\sqrt{1 - m^2}$. To portray how an inhomogeneous strain in $z$ direction can create pseudo-field in such a model we recall that strain modifies the hoppings according to (see e.g.~\cite{Pikulin2016})
\begin{eqnarray}
t_\parallel \sigma_z \to t_\parallel \sigma_z (1 - u_{zz}) + i t_\perp \sum_{j\neq z} u_{zj} \sigma_j.
\end{eqnarray}
Here $u$ is the symmetrized strain tensor $u_{ij} = \frac{1}{2}(\partial_i u_j + \partial_j u_i)$, where $\mathbf{u}$ is the displacement vector. Let us concentrate on the simplest case where $\mathbf{u} = (0, 0, \alpha z)$. Then $u_{33} = \alpha$ and the  separation between Weyl nodes is modified according to $|\mathbf{b}| \to |\mathbf{b}| - \alpha |\mathbf{b}| / (a |\mathbf{b}|)^2$. In the case of the position-dependent strain this results in the position-dependent Weyl node position and pseudofields.

Strain can induce more terms than those associated to pseudofields, which are are the main focus of this review. 
Ref.~\cite{Arjona2018} listed the richer structure that $u_{ij}$ and its antisymmetric counterpart $w_{ij}=\frac{1}{2}(\partial_i u_j - \partial_j u_i)$ can induce in the low energy theory. 
For example, $u_{ij}$ induces an anisotropic Fermi velocity~\cite{Cortijo2016} and its trace a deformation potential~\cite{Arjona2017} and the vector $A_i = w_{ij} b_j$ is responsible for a linear dispersion tilt and a pseudo-Zeeman term~\cite{Arjona2017}. 
The antisymmetric part of $w_{ij}=\epsilon_{ijk}\Omega_k$ with $\boldsymbol{\Omega} = \frac{1}{2}\nabla\times\mathbf{u}$ describes rotational strain and is independent of the separation of the Weyl nodes in momentum space. 
It also allows to tilt the dispersion, add a deformation potential, pseudo-Zeeman terms and pseudo-gauge fields~\cite{Arjona2018}, yet their physical consequences remain to be fully explored.

\section*{BOX 2: Fermi surface in the quantum limit in the presence of external and pseudo-magnetic fields}
\begin{figure}
    \centering
    \includegraphics[scale=0.25]{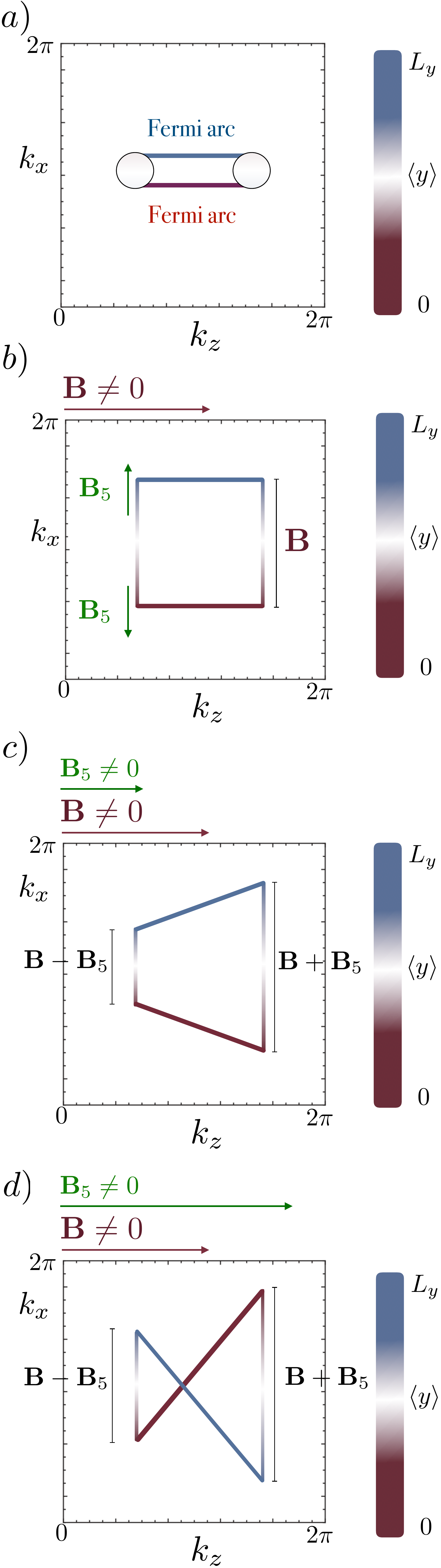}
\end{figure}

A finite sample of Weyl semimetal, say in the $y$ direction, has topologically protected Fermi arc states. This can be illustrated by iso-energetic contours for a finite Weyl semimetal slab of length $L_y$. See subfigure \textbf{(a)}.
These are localized at the two surfaces $y=0$ and $y=L_y$ (red and blue lines correspondingly) where $L_y$ is the total length of the finite slab. Bulk states are illustrated by the white circles

In the presence of a magnetic field, a finite sample of Weyl semimetal will feature Landau levels, dispersing in the direction of the magnetic field, and Fermi arcs, dispersing along the surface $\mathbf{B}_5$.
As worked out by Refs.~\cite{Bulmash2016,Ominato2016}, the Fermi surface thus traces a rectangle if a large magnetic field is along the Weyl node separation, as depicted in subfigure \textbf{(b)}, and a parallelogram for a generic field orientation.
This two observation can be traced back to position-momentum locking: the $k_x$ momentum is related to the average $\left\langle  y \right\rangle$ position of a bulk Landau level, while the additional two sides are the Fermi arcs, or equivalently, the pseudo-Landau levels for the surface $\textbf{B}_5$.

If we engineer a bulk pseudo-magnetic field, on top of the one that created the Fermi arcs, each chirality will feel an effective magnetic field, $\mathbf{B}_{\chi}=\mathbf{B}+\chi\mathbf{B}_5$. 
These will hence reduce and increase the Landau level degeneracy for negative and positive chiralities respectively, shrinking and elongating the sides of the parallelogram and making it a prism.
The resulting Fermi surface is depicted in subfigure \textbf{(c)}.

When the bulk pseudo-magnetic field increases beyond the value of the real magnetic field, the Fermi surface becomes a bow-tie, shown in subfigure \textbf{(d)}.

%\pagebreak

\begin{widetext}

\section*{BOX 3: Consistent and covariant anomalies:}

In this Box, we summarize the known subtleties of the chiral anomaly and its physical consequences in condensed matter (see Ref.~\cite{Landsteiner2016} for a review).
In the presence of electromagnetic fields, and importantly pseudo-electromagnetic fields, the number of chiral fermions is not independently conserved.
A way of arriving at this result is to consider a single Weyl node under the action of electromagnetic and axial electromagnetic fields.
From our discussion in the main text a Weyl node of chirality $\chi=\pm$ feels a total field that is given by a combination of real and axial fields $\mathbf{B}_{\chi}=\mathbf{B}+\chi\mathbf{B}_5$ and $\mathbf{E}_{\chi}=e\mathbf{E}+\chi\mathbf{E}_5$. 
The magnetic field creates Landau levels of degeneracy proportional to $\mathbf{B}_{\chi}$. 
Under the action of the field $\mathbf{E}_\chi$, the filled states are pumped at a rate proportional to $\mathbf{E}_\chi$.
Combining these two observations~\cite{Nielsen1981} the current conservation law for a given chirality can be written as~\cite{Bertlmann2000,Liu2013,Landsteiner2016}:
\begin{equation}
\label{eq:chiral anomaly}
    \partial_t \rho_\chi + \nabla \cdot \mathbf{J}_\chi =\dfrac{1}{8\pi^2} \mathbf{E}_\chi\cdot \mathbf{B}_\chi,
\end{equation}
where we chose natural units and $\rho_\chi$ and $\mathbf{J}_\chi$ are the density and current density of each chirality respectively.
Adding and subtracting the two chiralities will lead to conservation equations for both the total and chiral charges and currents respectively. 
Proceeding, leads to a surprise
\begin{subequations}
\label{eq:chiral anomaly}
\begin{eqnarray}
\label{eq:chiral anomaly1}
   \partial_t \rho_5 + \nabla \cdot \mathbf{J}_5 &=&\dfrac{1}{4\pi^2}(\mathbf{E}\cdot \mathbf{B}  + \mathbf{E}_5\cdot \mathbf{B}_5), \\
\label{eq:chiral anomaly2}
     \partial_t \rho + \nabla \cdot \mathbf{J} &=&\dfrac{1}{4\pi^2}(\mathbf{E}\cdot \mathbf{B}_5  + \mathbf{E}_5\cdot \mathbf{B}),
\end{eqnarray}
\end{subequations}
where $\rho_5$ ($\rho$) and $\mathbf{J}_5$ ($\mathbf{J}$) are chiral (total) charges and currents respectively, which we can combine in the four vectors $J^{\mu}=(\rho,\mathbf{J})$ and $J_5^{\mu}=(\rho_5,\mathbf{J}_5)$. 
Surprisingly, for these equations it appears that both the axial \textit{and} (more worryingly) the total charge are not conserved in the presence of both electromagnetic and axial electromagnetic fields.

Charge non-conservation in the presence of axial fields is an old conundrum, first encountered and solved in high-energy physics, with far reaching consequences for condensed matter. 
At the expense of (some) physical intuition we could redefine the total current such that the right hand side of Eq.~\eqref{eq:chiral anomaly2} is zero~\cite{Bardeen1984,Gorbar2017c,Gorbar2017a}.
The extra current terms that must be added to do so are known as Bardeen Polynomials in high-energy physics~\cite{Bardeen1984,Gorbar2017c,Gorbar2017a}.
A simple example for such added current is the Hall current proportional to the Weyl node separation $\mathbf{j}=\mathbf{b}\times \mathbf{E}$. 
Added to the current in Eq.~\eqref{eq:chiral anomaly2} it compensates exactly for the first term on the right hand side.
Similarly, other currents can be added to end up with 
\begin{subequations}
\label{eq:chiral anomaly}
\begin{eqnarray}
\label{eq:chiral anomalycons1}
   \partial_t \rho_{5,\mathrm{cons}} + \nabla \cdot \mathbf{J}_{5,\mathrm{cons}} &=&\dfrac{1}{4\pi^2}(\mathbf{E}\cdot \mathbf{B}  + \dfrac{1}{3}\mathbf{E}_5\cdot \mathbf{B}_5), \\
\label{eq:chiral anomalycons2}
     \partial_t \rho_\mathrm{cons} + \nabla \cdot \mathbf{J_\mathrm{cons}} &=&0,
\end{eqnarray}
\end{subequations}
known as the consistent version of the covariant anomaly equations, which conserve total charge.

Adding the Bardeen polynomials fixes the problem of charge conservation, yet they sacrifice physical intuition of how this exactly happens at a microscopic level, namely, where does the added piece of the current come from.
At a mathematical level the currents correcting the electric current and density are determined by the Berry curvature, and thus of topological origin~\cite{Gorbar2017a}.
In  contrast, the chiral current is corrected by a term which depends on lattice parameters and the definition of chirality~\cite{Gorbar2017l,Behrends2018}.

The different anomaly terms in the consistent and covariant pictures have concrete physical meanings in lattice systems~\cite{Gorbar2017a,Gorbar2017l,Behrends2018}.
To grasp their meaning it is helpful to understand how each term in the covariant version acts on the square and bow-tie Fermi surfaces discussed in Box 2.
In the figure in Box 2 we interpreted the Fermi arcs as pseudo-Landau levels.
The application of an external electric field along the direction of the corresponding $\textbf{B}_5$ pumps charge between different surfaces, by an amount proportional to $\textbf{E}\cdot\textbf{B}_5$.
This is compensated by a bulk Hall current, of magnitude $\mathbf{j}=\mathbf{b}\times \mathbf{E}$ which restores charge conservation.

The second term on the right hand side of Eq.~\eqref{eq:chiral anomaly2} can have a different interpretation. 
From a Fermi surface point of view the emergence of $\mathbf{E}_5$ is a time-dependent Weyl node separation. 
From a microscopic point of view, a rigid shift of the band bottom can precisely result in a shift of this kind if we focus our attention
to a region close to the Weyl nodes.
The apparent charge creation is not such if we consider the whole band, which is moving rigidly.
In this view, the second term in Eq.~\eqref{eq:chiral anomaly2} is an artifact from insisting in observing changes from a frame where we move with the chemical potential.

A notable difference between consistent and covariant pictures is the 1/3 prefactor between the second term on the right hand side of Eq.~\eqref{eq:chiral anomaly1} and its counter part \eqref{eq:chiral anomalycons1}. 
Its value is non-universal and depends on the pseudo-field configuration.~\cite{Gorbar2017l,Behrends2018}.

Putting all together, the covariant currents take contributions from carriers near the Fermi surface, while the consistent currents also add the information of the filled states, such as a finite Hall conductivity (see table).
\begin{figure}
\includegraphics[scale=0.2]{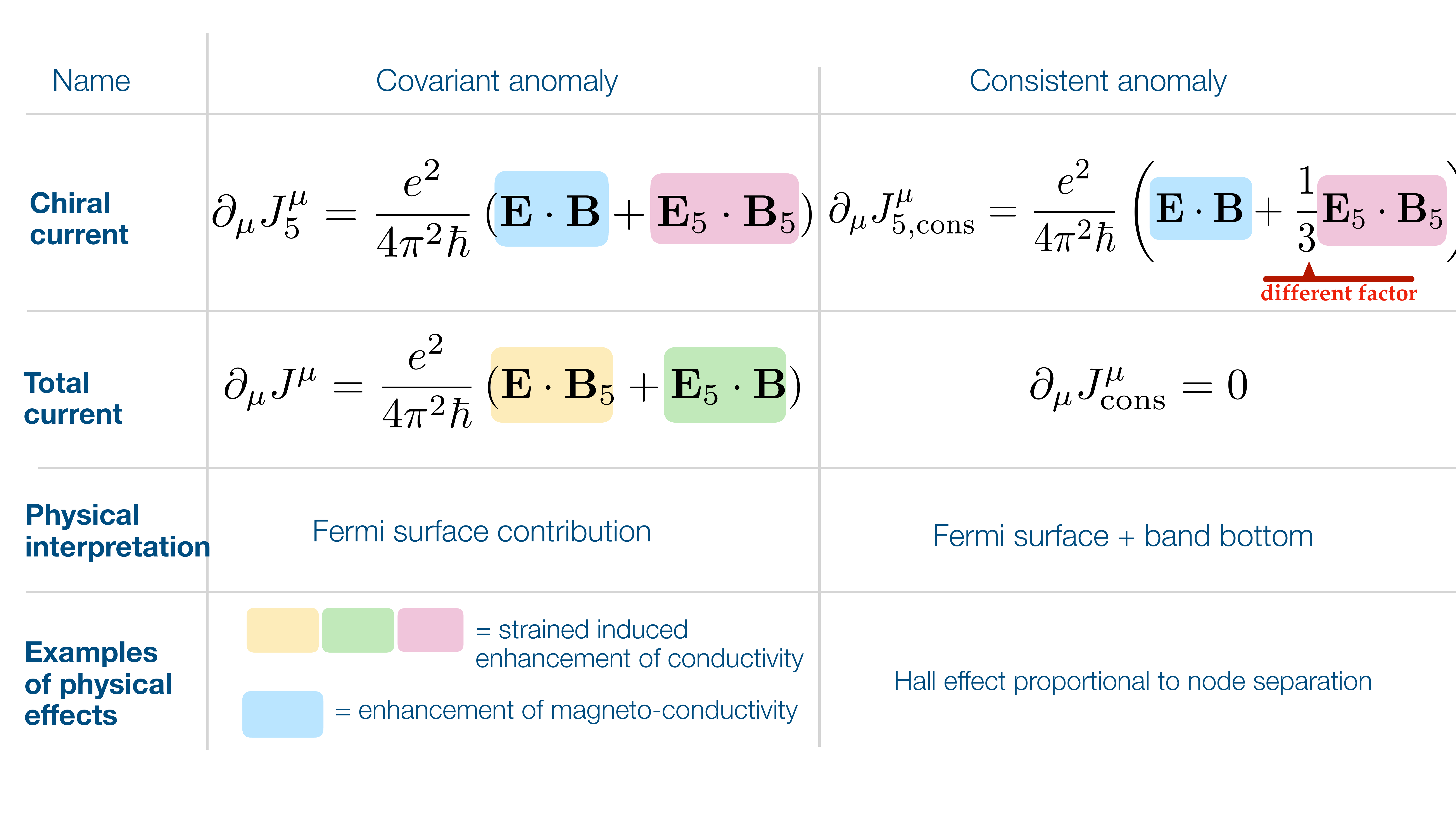}
\end{figure}
\end{widetext}

\end{document}